\documentclass[a4paper,11pt]{article}
\usepackage{generic,btp,times,url}

\InitEpsf{0.5}

\usepackage{amssymb}
\usepackage{amsfonts}
\usepackage{latexsym}
\usepackage{epsfig}

\newlength{\expandwidth}
\newlength{\expandheight}
\title{Business Suitability Principles\\
       for \\
       Workflow Modelling{\footnote{Part
of this work has been supported by CITEC, a business unit 
of the Queensland Government's Department of Public Works 
and Housing (formerly the Administrative Services
Department).}}}

\author{A.P. Barros${^1}$,
        A.H.M. ter Hofstede$^1$,
        H.A. Proper${^2}$,
        P.N. Creasy${^1}$ \\[0.3cm]
        \begin{tabular}[t]{c@{\extracolsep{2em}}c}
        \small $^1$Department of Computer Science &
        \small $^2$Faculty of Information Technology \\
        \small The University of Queensland &
        \small Queensland University of Technology \\
        \small Brisbane, QLD 4072 &
        \small GPO Box 2434, Brisbane QLD 4001 \\
        \small Australia &
        \small Australia \\
        \small & e-mail: E.Proper@acm.org
        \end{tabular}
        }
\date{}

\begin{document}

\maketitle

\begin{quote}
\begin{tabbing}
                \> Object-Oriented Analysis
\end{tabbing}
\end{quote}

{\sc Published as:}
\begin{quote}
  A.P. {Barros}, A.H.M.~ter {Hofstede}, H.A.~(Erik) {Proper}, and P.N. {Creasy}. {Business Suitability Principles for Workflow Modelling}. Technical report, Department of Computer Science, University of Queensland, Brisbane, Queensland, Australia, August 1996.
\end{quote}

\begin{abstract}
By incorporating aspects of coordination and collaboration, workflow
implementations of information systems require a sound conceptualisation
of \EM{business processing} semantics. Traditionally, the success of 
conceptual modelling techniques has depended largely on the adequacy of 
conceptualisation, expressive power, comprehensibility and formal foundation. 
An equally important requirement, particularly with the increased 
conceptualisation 
of business aspects, is \EM{business suitability}. 

In this paper, the focus is on the business suitability of workflow modelling 
for a commonly encountered class of (operational) business processing, 
e.g. those of insurance claims, bank loans and land conveyancing. A general 
assessment is first conducted on some \EM{integrated} techniques characterising 
well-known paradigms - structured process modelling, object-oriented modelling, 
behavioural process modelling and business-oriented modelling. Through this, an 
insight into business suitability within the broader perspective of technique 
adequacy, is gained.
A specific business suitability diagnosis then follows using a particular 
characterisation of business processing, i.e.\ one where the intuitive 
semantics and inter-relationship of business services and business processes 
are nuanced. As a result, five business suitability principles are elicited. 
These are proposed for a more detailed understanding and (synthetic) development 
of workflow modelling techniques. Accordingly, further insight into workflow 
specification languages and workflow globalisation in open distributed architectures 
may also be gained.
\end{abstract}
\section{Introduction}
\label{intro}

The workflow concept, proliferated through the recently emergent
computer supported cooperative work (CSCW) systems and workflow
systems (see surveys in 
\cite{Report:94:Fitzpatrick:CSCW,%
Book:93:Wastell:CSCW,Article:91:Rodden:CSCW}
and
\cite{Article:95:Georgakopoulos:WFOverview} respectively), 
advances information systems (IS) implementation models by incorporating 
aspects of collaboration and coordination in business processes. Under 
traditional implementation models, applications are partitioned into 
discrete units of functionality, with (typically) operational procedures 
used to describe how human and computerised actions of business processes 
combine to deliver business services. Through an endowment of business 
process execution semantics, workflows permit a greater organisational 
fit of ISs. Moreover workflows are specified at a level above traditional 
applications, enabling program binding and access to a loosely-coupled set 
of databases and files. Therefore, newer applications may be developed out 
of existing applications to reflect reengineered business processes. 

Crucial to the specification of any IS implementation is the \EM{conceptual}
level. This, of course, orients the analysis of a given domain towards its
essence (deep-structure) rather than to aspects of implementation
(physical-structure) or representation (surface-structure). 
It is a well-known fact the later problems and
inadequacies are detected in specifications, the greater the expense of
correction
\cite{Book:90:Davis:SoftReq}. 

For workflows, the standardisation of concepts is progressing 
through the Workflow Management 
Coalition{\footnote{Refer to http://www.aiai.ed.ac.uk/WfMC/index.html for
more details.}}.
While the set of terms and references defined so far characterise
sufficiently the notion of workflow, e.g.\ event, process (including
pre-conditions, post-conditions and state transitions) and organisational
(or actor) role, much of the focus is geared towards workflow management
systems and their specification languages. The emphasis is on part of
business processing, namely process execution semantics: sequence, repetition,
choice, parallelism and synchronisation. A sound conceptualisation requires
not only this but also that process semantics, e.g.\ the messaging, database
updates and retrievals involved, to be explicitly captured.

In general, for the conceptual level, \EM{techniques} are available
under different paradigms, for the modelling of different
aspects of a business domain (see e.g.\
\cite{Book:88:Olle:Methodologies}). When integrated into well-formed
methods, integrated IS specifications - result.
A number of paradigms may be discerned for workflow modelling: 
process-centric, e.g.\
\cite{Article:91:Dur:TaskActor};
state-centric, e.g.\
\cite{Article:95:Antonellis:OOA};
and actor-centric, e.g.\
\cite{Article:94:Dietz:TransModelling}
(based on the speech-act theory synthesis of
\cite{Article:80:Flores:ComMod}).
Moreover the use of business (or enterprise)
models, e.g.\ as deployed in requirements engineering methods
\cite{Article:95:Berztiss:ReqEng,%
Article:95:Loucopoulos:ReqEng,%
Article:94:Anton:ReqEng},
in design methods
\cite{PhdThesis:93:Ramackers:BehaviourMod}
and in CAiSE tools e.g.\ AD/CYCLE
\cite{Article:90:Mercurio:CAiSE},
provides an \EM{organisational embedding} whereby a workflow model's
components may be backtracked to its real-world counterparts.

Although, the field of conceptual modelling has become fairly mature, the 
application of techniques, has, by and large, followed the intuition of the 
developers of models. This, of course, involves an informal to formal 
transition. 
With workflow specifications, this transition is reduced, however a greater 
alignment is required between the workflow modelling cognition and business 
processing cognition. Beyond the qualification of fundamental modelling 
concepts 
(e.g.\ process) with organisational attributes (e.g.\ business service), the 
business 
processing semantics need to be infused into the semantics of a technique such 
that a workflow may be expressed and communicated adequately using that 
technique. 
In absence of a universal organisational theory, much uncertainty exists as to 
how effective conceptual modelling techniques are for business workflows; 
whether, 
given the diversity of business processing, any generable prescription of 
business
processing cognition is in fact possible or desirable.

This paper addresses some of this uncertainty. It recognises that there are
a basic set of requirements which conceptual modelling techniques should
fulfill in order to be effective. These requirements are based on the 
well-known conceptual modelling principles of
\cite{Report:82:ISO:Concepts}. If not properly catered for, the development
of workflow specifications may be problematic.

First and foremost, in accordance with the \EM{Conceptualisation Principle},
a technique should focus on essential detail only. 
The focal aspects of business models and workflow specification languages can 
deflect and sometimes dictate the quality of conceptualisation. Conversely, 
following the \EM{One Hundred Percent Principle}, techniques should provide 
a sufficient \EM{expressive power} so that a full conceptualisation is in fact 
possible. If the expressiveness of a workflow specification language exceeds 
that of a technique,  where only a partial conceptualisation results, the 
remainder has to be addressed at the implementation level.
This problem is known in software engineering jargon 
as the \EM{waterfall}. Such a disjointness of analysis precludes a sound 
understanding of the problem, leading to premature implementations. 
Conceptual models also need to be communicated and validated with 
a diverse set of stakeholders and so mechanisms are required for an 
effective \EM{comprehensibility}. An effective graphical presentation 
together with abstraction/decomposition mechanisms facilitates this.
At the same time, a \EM{formal foundation} is required to 
prevent interpretation ambiguities and to enable formal reasoning. That 
a recent survey on workflow technology
\cite{Article:95:Georgakopoulos:WFOverview}
cited the problem that ``workflow models and process methodologies do not
explicitly support the specification of what it means for a workflow to
be correct'' is symptomatic of a lack of formal semantics. Together with a
formal syntax, this constitutes formal foundation. 
Finally and most relatedly to the issue of the paper, since a 
``silver bullet'' 
for all types of domains is considered unrealistic (see e.g.\
\cite{Article:87:Benyon:SystemAnalyst,%
Article:87:Brooks:SilverBullet,%
Article:83:Malouin:ISMethods}),
techniques should be \EM{suitable} for their problem domains. This means
a close connection between the modelling concepts and features, and those 
required by the domain.

Of significance for business processing domains is the 
\EM{business suitability} 
of techniques. Since there are many types of organisations and many types of 
business processing 
\cite{Book:85:Davis:MIS},
particular attention is drawn to that type of (operational)
business processing which exhibits precise execution paths. As examples, 
the processing of insurance claims, bank loans and land conveyancing, are 
mission-critical in nature and are rarely undertaken without strict 
operational procedure. 
Also, multiple interactions with clients and external organisations are
typically needed to fulfill service requests.
Given their closer connection with database technology, workflow systems 
are considered more appropriate for this type of business processing over 
CSCW
\cite{Article:92:McCready:WFS,Article:95:Georgakopoulos:WFOverview}. 
These notions of business processing, workflow and workflow systems are 
assumed hereafter.
 
Following the constructivist approach, 
notably as advocated for IS concept development in
\cite{Report:96:Falkenberg:ConcFramework},
a number of \EM{integrated} conceptual modelling techniques 
- are assessed using the aforementioned requirements.
Attention then turns to determining common problems of business suitability, 
and in 
turn, the formulation of a number of principles.
The result is five principles. The \EM{\Embed\ Principle} describes how a model
should be backtracked to organisational elements. The \EM{\Validate\ Principle}
identifies the need for scenarios, and in particular, a business transaction,
as distinct from business process and business service, for workflow cognition.
The \EM{\InfoHiding\ Principle} requires that business processing undertaken
for business service requests should be insulated from the requests, and in
so doing, motivates the need for an explicit treatment of business services
within conceptual modelling. The
\EM{\Combine\ Principle} requires that all concepts involved in workflow model
enactment be simultaneously present in the model. Simultaneously absent in the
assessed techniques were the combination of structural \EM{and} behavioural 
aspects of workflows, human to computer interaction and temporal aspects. 
Finally, the \EM{\ErrorHandling\ Principle} identifies the need for 
operational 
error handling to be catered for at the conceptual level, thereby 
incorporating 
the recovery management focus of \EM{transactional} workflows into a general
exception handling. An improved insight into the assessment and development of 
workflow modelling techniques, is claimed. Also the proposed separation of 
business service, business process and business transaction, can lead to a 
more 
effective globalisation of workflows, where, for example, within open 
distributed
environments (e.g. the Web), services involving workflows can be ``mixed 
and matched''.

The paper is organised as follows. In section~\ref{survey}, techniques 
characteristic of the different paradigms - structured
process modelling, object-oriented modelling, behavioural process modelling
and business-oriented modelling - are assessed. In 
section~\ref{new-principles},
the problems and business suitability principles are defined. In 
section~\ref{epilogue}, the paper is concludes with an epilogue.

\section{An assessment of integrated conceptual modelling techniques}
\label{survey}

In this section, an insight into the capabilities of techniques to support
a sound conceptualisation of business processing for workflow specifications
is sought.
For this, an assessment of some integrated conceptual modelling 
techniques is conducted using the general conceptual modelling requirements
(described in section~\ref{intro}). These are: conceptualisation,
expressive power, formal foundation, comprehensibility and business
suitability. The techniques are drawn from the well recognised paradigms 
of:

\begin{itemize}
   \item structured process modelling (section~\ref{str-process})
   \item object-oriented modelling (section~\ref{oo})
   \item behavioural process modelling (section~\ref{beh-process}) 
   \item business-oriented modelling (section~\ref{bus-model})
\end{itemize}

The approach to the assessment is qualitative, reflecting its
motivation. For this, the key areas of observation include
integration strategy, integration structure and the mechanisms used to 
adapt concepts and features for the business level. An integration strategy 
reflects the
cognitive dependency of a technique's partial models for an overall 
understanding 
of a business domain. Process modelling techniques centralise processes with 
respect to data flow, control flow and data repositories, and so 
a \EM{process-centric} integration with data models follows. The dual of this 
is a \EM{state-centric} integration, where object modelling
techniques centralise object states with respect to processes involved in state
transitions, and so process modelling follows.

An integration structure relates to the definition of concepts in the
partial models. In a \EM{tightly-coupled} structure, all concepts are
bound to a single logical definition. In a \EM{loosely-coupled} structure, 
each partial model has a separate definition, and some concepts may be 
integrated for basic consistency. Clearly, a tight-coupling permits a stronger 
expressive power and formal foundation. However, a loose-coupling provides 
greater flexibility - particularly for ``contingency frameworks'' 
\cite{Article:91:Avison:InfSystResearch}
- where 
different techniques may be selected depending on the analysis ``situation''. 
Since a (relatively) specific business processing situation underscores this 
paper, the former is preferred in the assessment.  

\subsection{Structured Process Modelling}
\label{str-process}

Structured process modelling, for example Structured Analysis
\cite{Book:86:Gane:SSAD,Book:89:Yourdon:Analysis,%
Book:78:DeMarco:StructAnalysis}
and ISAC \cite{Book:81:Lundeberg:ISAC} 
have had widespread use, providing both top-down process 
analysis and software design mapping. With the prevalence of database
technology, a number of extensions have been proposed, e.g.
\cite{Article:91:Shoval:IntegDesign,%
Book:89:Yourdon:Analysis},
to incorporate data modelling techniques. In particular, process models 
are often used for the higher levels of analysis, and the identified data 
repositories lead to data modelling; clearly a process-centric
integration strategy.

Data Flow Diagrams (DFD), associated with Structured Analysis, are 
a popular - arguably the most popular - structured process modelling 
technique. A DFD is a directed graph where the nodes represent external 
entities, processes and data stores, and where the edges represent data 
flows between the nodes. A process may be decomposed into another DFD, provided 
the DFD contains at least that process's data flows. At the lowest level, each 
process has a detailed Structured English specification. Specifications for data 
flows and stores are contained in a data dictionary. 

\begin{figure}[htb]
\centering
\epsfig{figure=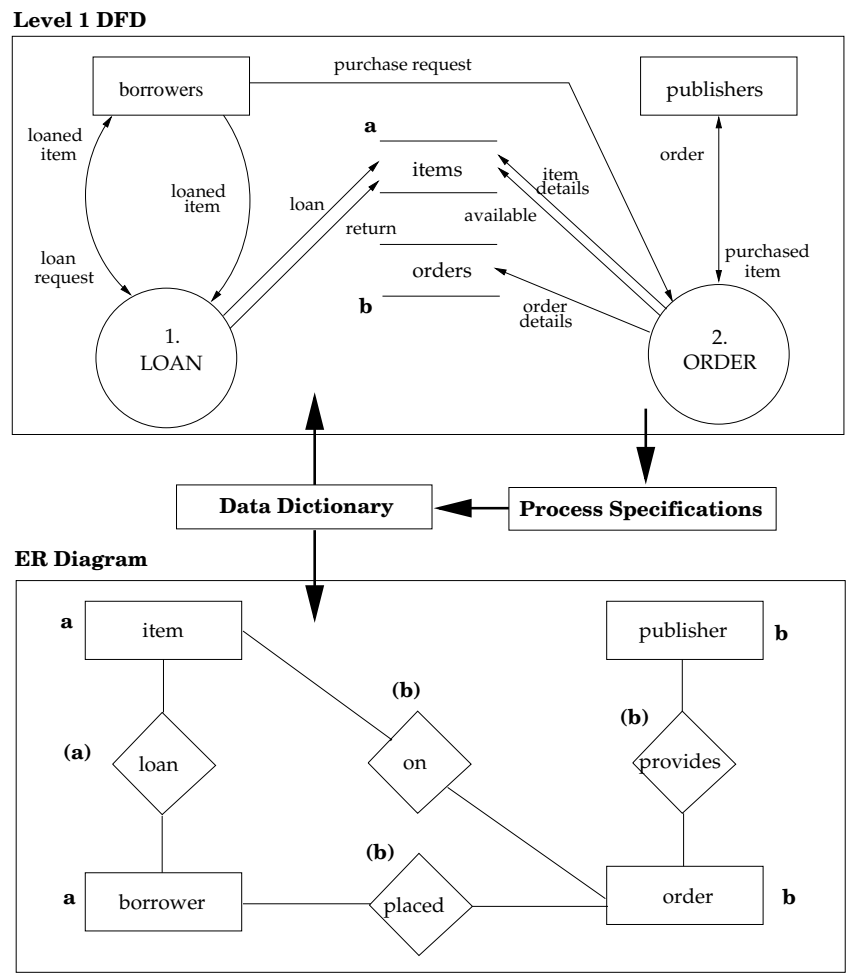,width=0.7\linewidth}
\caption{Process and data model integration for the library domain in 
Structured Analysis}
\label{yourdon.fig}
\end{figure}

Figure~\ref{yourdon.fig} illustrates a process-centric integration of process
and data models for a library domain using
\cite{Book:89:Yourdon:Analysis}. 
The data model is based on an Entity Relationship Model (ERM) technique. For 
brevity, the data dictionary and process specifications have been omitted.
 
The DFD models the processing of \textsf{loan requests} and \textsf{orders}, 
viz: \textsf{borrowers} (an external entity) send \textsf{loan requests} 
(data flow) for \textsf{items} to \textsf{LOAN} (process). A \textsf{loan} 
(an update) is issued in \textsf{items} (data store) and the \textsf{item} is 
passed to the \textsf{borrower}. When the \textsf{loaned item} is returned 
to \textsf{LOAN}, a \textsf{return} is issued to the \textsf{items} data 
store to indicate that the \textsf{item} is available for loaning again. 
\textsf{Borrowers} also make \textsf{purchase requests} (for an item) to 
the \textsf{ORDER} process. For this, \textsf{order details} are stored 
in \textsf{orders}, and \textsf{item details} are stored in \textsf{items}. 
An \textsf{order} is sent to a \textsf{publisher} who supplies the 
\textsf{ordered item}. Once supplied, the \textsf{item} is updates to be
\textsf{available}.

The data dictionary provides the integration between DFDs and ERMs. This is
achieved using syntactic correspondences. A correspondence is required between 
data stores and entity types, indicated by alphabets in Figure~\ref{yourdon.fig}. 
The integration structure is therefore loosely-coupled.
Furthermore, it is clear that the degree of integration is 
coarse. For instance, no correspondence exists for relationship 
types, unless they are aggregated. Sometimes processes, data flows and data
stores indicate relationship types as indicated by \textsf{(a)} and 
\textsf{(b)}. However, the absence of formal semantics - an omission
widely observed in most DFD techniques (see e.g.
\cite{Report:89:Bruza:DataFlowSem,%
Article:93:Opdahl:DFDs}
- makes a detailed correspondence difficult. Care should also be taken when
directly corresponding multi-object data stores since unnaturally 
aggregated entity types may result. 
One methodological implication is to delay ERM modelling until sufficiently 
detailed DFDs are drawn. Alternatively, 
\cite{Book:92:Batini:ConceptualDBDesign} 
proposes a joint methodology which uses mutually influential decompositions
in both ERMs and DFDs.

Process specifications utilise data dictionary elements to maintain
consistency with DFDs and ERMs. However, the expressive power of
Structured English is restricted by the basic level of integration.
Business rules including those expressed through object lifecycles are
implicit in process specifications. Process dependencies are determined 
in an ad-hoc manner solely for the purpose of systems design.

Clearly the strength of a DFD lies in its simple and general concepts.
Together with the decomposition feature, this allows an effective
comprehensibility, particularly for the early phases of analysis where 
the broader context of a system and its scope needs to understood. 
Additionally, a large DFD may be partitioned by external events, permitting 
horizontal views for validation. Like most classical process modelling 
techniques, DFDs may be applied to several domains including real-time 
processing and text processing. For business domains, its methodology 
recommends the modelling of organisational processing structure first, 
from which essential DFDs are determined. In this regard, its simplicity and 
generality can compromise the business suitability of the technique, 
e.g. the relationship with organisational processing structure is arbitrary 
and only basic aspects of business processes are modelled. Moreover,
the procedural style of expression for DFDs and process specifications
imposes an imperative rather than a declarative conceptualisation.  
At lower levels of abstraction, DFDs can prescribe implementation, 
thereby compromising conceptualisation. Further problems of DFDs are
cited in
\cite{Article:93:Opdahl:DFDs}.

\subsection{Object-oriented modelling}
\label{oo}

An object-oriented approach to programming, and in recent years, to analysis 
and design (OOA/D), stems from the premises that objects provide a closer 
semblance with reality and are less prone to change than processes. Along with 
static properties, objects include in their classification, dynamic 
properties - operations, implemented as methods on classes. In keeping with 
tightly-coupled object models, conventional OOA/D techniques,
\cite{Book:88:Shlaer:OO,%
Book:91:Booch:OOD-applications,%
Book:90:Coad:OO,%
Book:91:Rumbaugh:OO,%
Book:92:Jacobson:Use-Cases}
advocate principally a state-centric integration of conceptual models. 
Object structure is modelled using techniques adapted, for the most part, from
data modelling techniques, e.g.
ERM based \cite{Report:90:Engels:ConcMod}
and Object-Role Model (ORM) based
\cite{Report:95:Proper:CDMKernel}.
Object behaviour is modelled using techniques based 
on formalisms like finite 
state machines (FSM) 
\cite{Book:79:Hopcroft:FSM}
and Petri nets 
\cite{Book:81:Peterson:PetriNets,%
Book:85:Reisig:IntroPetriNets}.

Since object behaviour models describe the intra-object dynamics, a higher
modelling context is required in domains where object interaction is
significant. Process modelling and hence a process-centric integration 
can be useful for this. For example, DFDs are used in many OOA/D techniques 
including
\cite{Article:91:Shumate:Str-OO,%
Article:91:Solsi:Str-OO,%
Article:90:Sully:Str-OO},
despite the paradigmal differences which are sometimes regarded as incompatible
\cite{Article:95:Embley:OOACompare,%
Article:91:Firesmith:Str-OO}.
Alternatively, other mechanisms, e.g.\ use cases or scenarios
as in
\cite{Article:94:Pollacia:OFD,%
Book:92:Jacobson:Use-Cases},
have been developed to preserve a high-level state-centric context.

\begin{figure}[htb]
\centering
\epsfig{figure=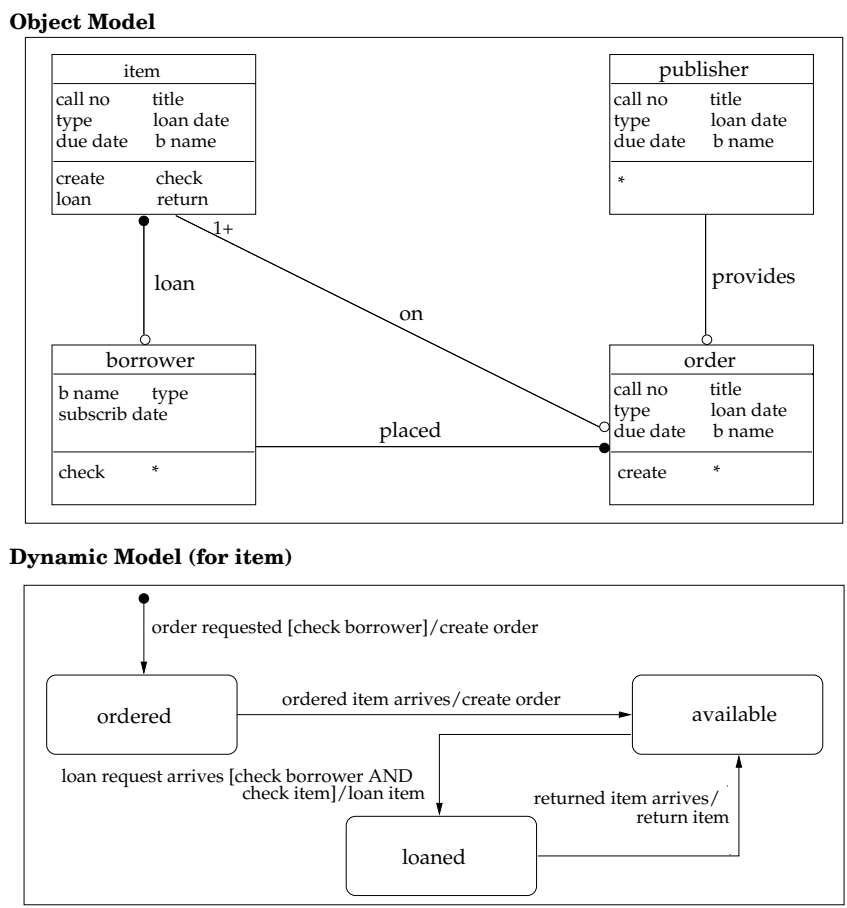,width=0.7\linewidth}
\caption{Object and dynamic model integration for the library domain in OMT}
\label{rumbaugh.olc.fig}
\end{figure}

A popular OOA/D technique, Object Modelling Technique (OMT)
\cite{Book:91:Rumbaugh:OO},
 incorporates both a state- and process-centric integration strategy. 
It accommodates the data, process and behaviour perspectives through 
an object model (ERM), a function model (DFD) and a dynamic model (FSM 
which includes nesting), respectively. Figure~\ref{rumbaugh.olc.fig} 
illustrates a state-centric model integration for the library domain.

In the object model, attributes are listed in the middle portion of 
the object class diagrams and methods are listed in the lower portion. 
Each object type has a dynamic model associated with it. The dynamic
model contains a partial order of behaviour states as nodes. States 
are defined by ``exclusive'' predicates and are orthogonal to the object 
type specialisations (a structural classification). The edges represent
state transitions. A state transition embodies an event which triggers
an action if a condition (optional) is satisfied. The significant result
of the action is, of course, the change to the next state. 

In Figure~\ref{rumbaugh.olc.fig} for example, the event \textsf{order requested} 
triggers the action \textsf{create order} to create an \textsf{order} object
(having no previous state) in the state \textsf{ordered} if some check(s) 
on borrowers are satisfied{\footnote{Textual aliases, in this case 
\textsf{check borrower}, have been used to denote condition predicates.}}.
A large class of business rules may be captured using state transitions. In fact,
event-condition-action (ECA) paradigm for rule-based languages is adopted in
active database technology, e.g. in context with object-oriented databases
\cite{Article:90:Chakravarthy:ActDB},
and object-oriented conceptual specification languages, e.g. as defined in the 
ESPRIT{\footnote{ESPRIT is an acronym for European 
Strategic Program for Research in Information Technology.}} project TEMPORA 
\cite{Article:91:Loucopoulos:Tempora}. 

\begin{figure}[htb]
\centering
\epsfig{figure=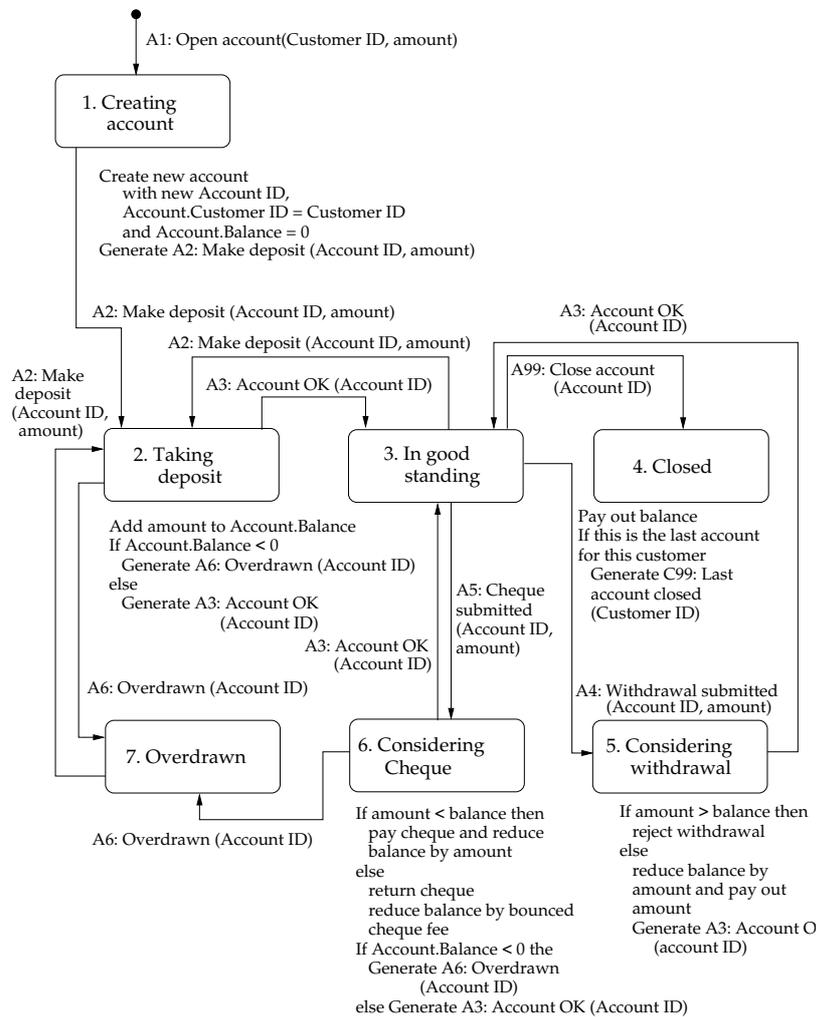,width=0.7\linewidth}
\caption{Object lifecycle for \textsf{accounts} of the bank domain in 
Shlaer \& Mellor}
\label{shlaer.fig}
\end{figure}

Digressing for a moment to another example,
Figure~\ref{shlaer.fig}, taken from 
\cite{Book:88:Shlaer:OO} (pp. 40),
illustrates a more complicated object lifecycle. An interesting feature of 
this example is the primary importance of state-centric integration for the 
domain. That is, the object behaviour model is pivotal to understanding the 
broader context of the domain. The pseudo code for state-transitions includes 
alternatives (illustrated) and iterations (not illustrated), improving the 
expressive power over ECA languages. In general, without a conceptual
specification language, techniques either omit detailed specifications or
convolute the graphical aspect of the model.  

\begin{figure}[htb]
\centering
\epsfig{figure=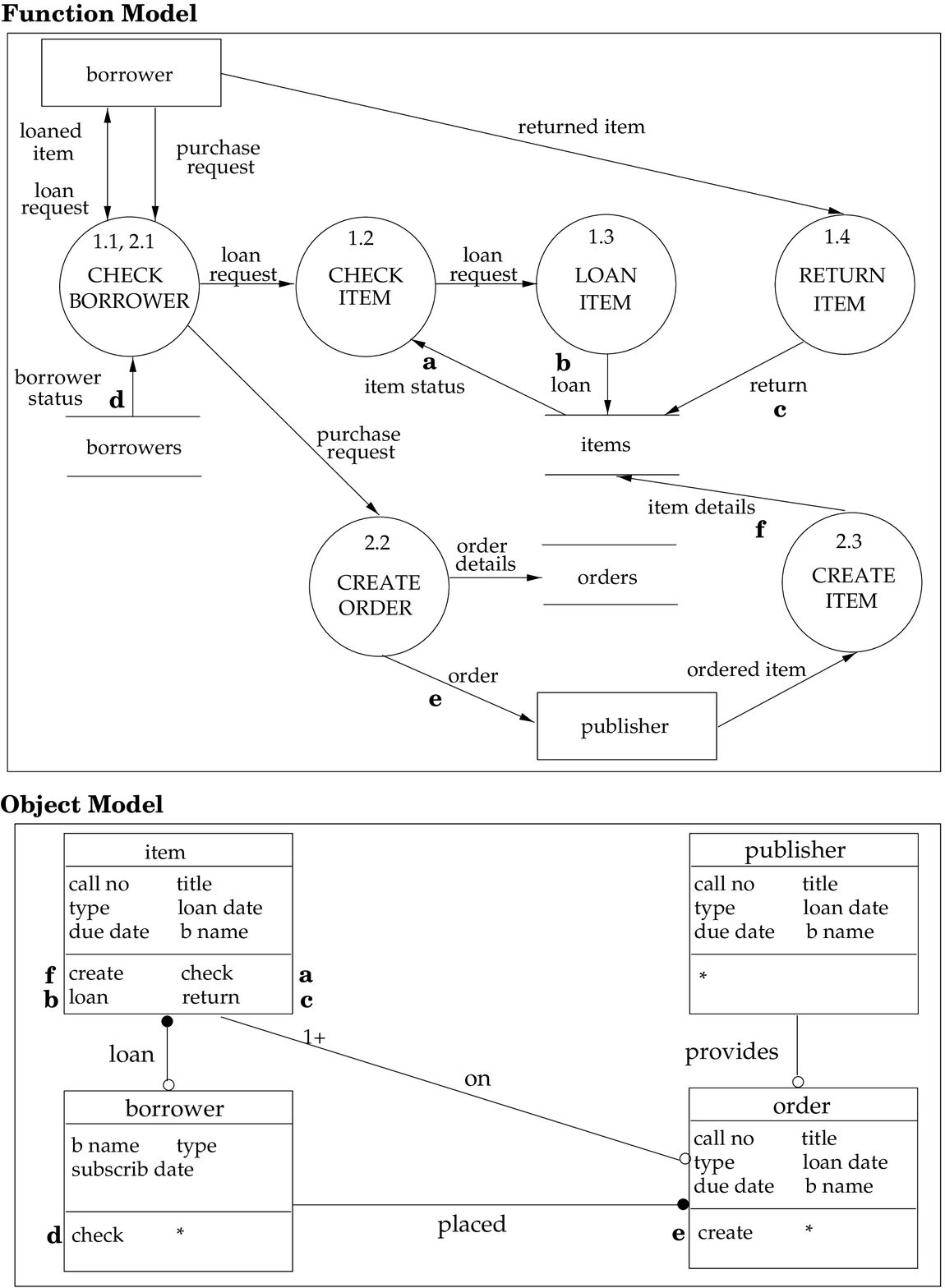,width=0.7\linewidth}
\caption{Function and object model integration for the library domain in OMT}
\label{rumbaugh.fig}
\end{figure}

Returning to OMT, Figure~\ref{rumbaugh.fig} illustrates a process-centric 
integration of the function and object models for the library domain.
Process-centric integration rules require that the function model's leaf 
processes correspond to object class methods. This is not straightforward since
a single leaf process may involve more than one object type. Clearly, the principal
object type referred to in OMT as the target, needs to be identified.
Targets are contained in data flows, data stores and actors 
(represented by a DFD's external entity symbol). A target may be identified 
by determining a ``client-server'' relationship between the object 
types. An object type is a target if it invokes requests from 
other object types for some purpose related to none other of those object
types. Hence the (lettered) correspondences in Figure~\ref{rumbaugh.fig};
a loosely-coupled integration structure.

Like other classical OOA/D techniques, OMT provides a simple and general set 
of concepts for several types of domains. For business domains, comprehensibility 
and suitability gains of the dynamic model are useful on a per object basis. The 
function model addresses the ``bigger picture'', however process control can only
be derived from the collective set of dynamic models. Despite claims of the 
``natural'' occurrence of objects, the technique's generality can present 
ambiguities in the analysis of organisational processing structures. For 
detailed specifications, a combined cognition of the models can help resolve 
difficult design decisions. For example, the function model provides further insight 
into the aggregation of complex object types, while the dynamic and function models
provide a collective insight into method identification. The 
lack of formal semantics (particularly for the function model) and expressive 
power restricts the effectiveness of these gains. In general, no evidence of formal 
semantics was found in the referenced OOA/D techniques.

\subsection{Behavioural process modelling}
\label {beh-process}

The enhanced suitability which results from behavioural (or control) 
aspects of process models has already been identified in the preceding
sections. Integrated techniques such as IML Inscribed Petri Nets
\cite{Article:82:Richter:IML} 
and Activity-Behaviour Modelling
\cite{Article:86:Solvberg:BehMod} 
pioneered the inclusion of behavioural process modelling in at least one 
level of process model decomposition. Behavioural aspects include process sequence, 
repetition, parallelism and synchronisation. Accordingly, collaborative aspects 
of processes can be modelled and problems such as deadlock, livelock
and starvation may be determined prior to implementation.

Classical Petri nets, Condition/Event-nets and Place/Transition-nets, have been 
adapted for process behavioural specification 
\cite{Article:81:Antonellis:ModelEvents}
but are limited in the degree of formal interpretation and real-world 
practicality
\cite{Article:91:He:PrTNets}.
High-level Petri nets such as Predicate/Transition nets (PrT-nets) 
\cite{Article:87:Genrich:ActivPrTnets}
and
Coloured Petri nets (CP-nets) 
\cite{Article:91:Jensen:ColouredPetriNets} 
were proposed to overcome these problems by the provision of declarations 
and net inscriptions, expressed in a formal language (based on first-order 
predicate logic for PrT-nets and a functional programming language for CP-nets). The 
incorporation of formal semantics leads to executable process specifications, 
e.g.\ in tools such as ExSpect 
\cite{Article:89:Hee:Exspect}
(hierarchical CP-nets) and Income/Star 
\cite{Article:93:Jaeschke:IntegTool} 
(Fuzzy Nets). 
\cite{PhdThesis:93:Verkoulen:ExSpect} formally demonstrates an 
integration of an object-oriented modelling technique, Simcon, and PrT-nets. 
The combination of both state-centric and process-centric constructs
into a single model, however, makes even high-level Petri nets difficult 
to comprehend.  Different strategies are used to maintain comprehensibility
without losing expressive power. 

\begin{figure}[htb]
\centering
\epsfig{figure=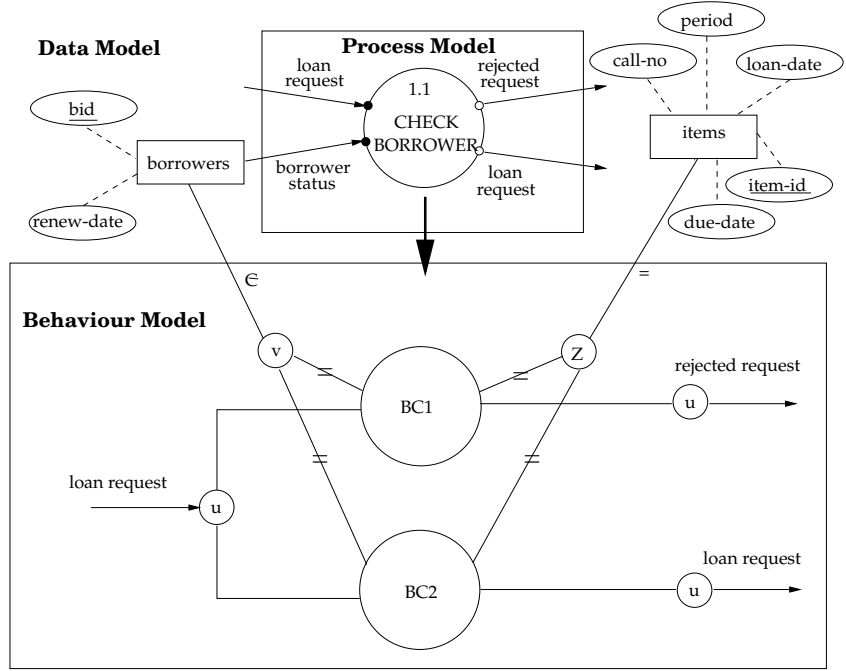,width=0.7\linewidth}
\caption{Data, process and behaviour model integration for the 
library domain in BNM}
\label{kung.fig}
\end{figure}

The Behaviour Network Model (BNM) 
\cite{Article:93:Kung:BNM} 
transfers behavioural aspects from higher levels of abstraction into the lowest 
level. In it, DFDs are used at higher levels, and at the lowest level, each process 
is transformed into a PrT-net which is tightly-coupled with an ER schema. Thus,
PrT-net specifications replace traditional Structured English. At 
all levels of abstraction, model integration is process-centric. As an example, 
Figure~\ref{kung.fig} illustrates a DFD, PrT-net and ERM integration for the 
library's \textsf{CHECK BORROWER} process.
 
The DFD specifies that the conjunction (black dot) of \textsf{loan request} 
and \textsf{borrower status} are required by \textsf{CHECK BORROWER} so that 
a disjunction (white dot) of either \textsf{loan request} or
\textsf{rejected request}{\footnote {This data flow has been introduced to
allow for model execution.}} is produced. Data flows are message carriers
which may be associated with ERM object types.
In a PrT-net, input flows, in this case \textsf{loan request}, connect to 
places (i.e.\ \textsf{u}, \textsf{v} and \textsf{Z}) 
which in turn connect to transitions (i.e.\ \textsf{BC1} and \textsf{BC2}).
Individual (element-of sign) or sets (equality sign) 
of instances of ERM entity types are linked to PrT-nets via places. 
Transitions, representing elementary actions, use 
entity and data flow instances as operands in pre- and post-condition rules. 

In the example, two rules are specified for \textsf{CHECK BORROWER}; note
the declarative rule specification possible through Petri nets compared to 
the imperative approach of Structured English.
Firstly \textsf{BC1}, that a \textsf{loan request} becomes a \textsf{rejected
request} if either the \textsf{borrower} has an overdue \textsf{item} or
will not be registered during the loan period: 
\begin{eqnarray*}
pre & & \neg \Ex{z \in Z}{u.\mbox{bid}=z.\mbox{bid} \land z.\mbox{loan-date}>
\mbox{\$TODAY}}
\lor \\
& & u.\mbox{bid}=v.\mbox{bid} \land v.\mbox{renew-date}<\mbox{DATE}(\mbox{\$TODAY,}
v.\mbox{period}) \\
post & & u.\mbox{msg}=\mbox{`Loan request rejected'}
\end{eqnarray*}

and secondly \textsf{BC2}, to ensure that a \textsf{loan request} is accepted 
(the negation of BC1). The specification for this is not provided. Note, the 
variable \$TODAY and the function DATE are introduced.

Compared to Structured Analysis, BNM's DFD technique is as suitable, though 
improved in comprehensibility. The use of PrT-nets for the conceptualisation of 
process specifications improves expressive power and provides a basis for formal 
foundation. Process pre- and post-conditions allow a further range of business
rules to be specified compared to object lifecycles. Object lifecycles however,
are implicit in one or more PrT-nets, and so the related class of business rules 
are specified implicitly.

\begin{figure}[htb]
\centering
\epsfig{figure=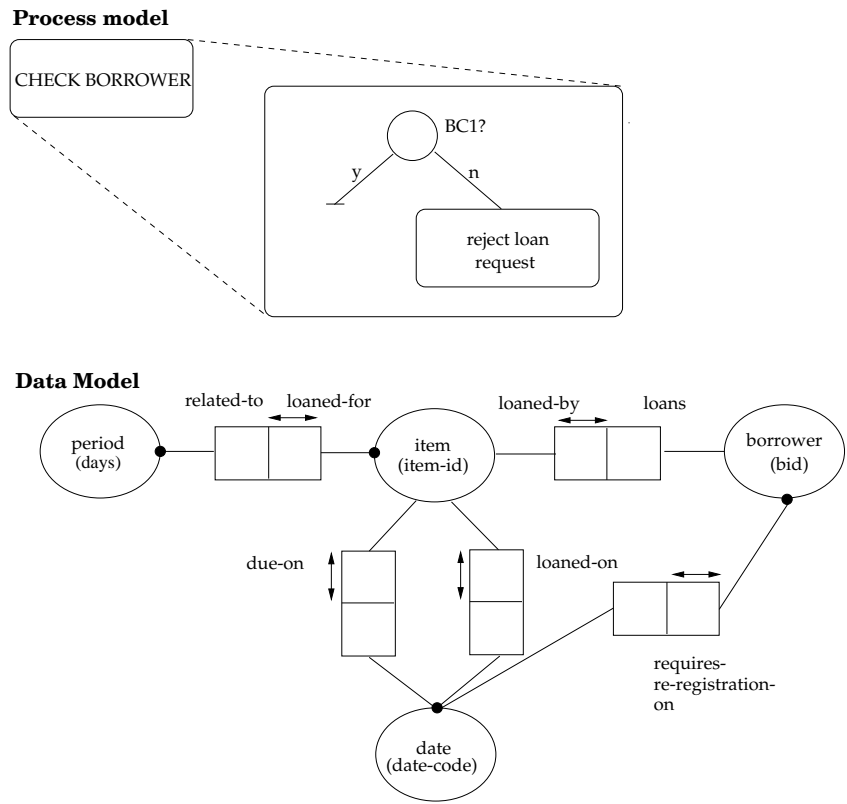,width=0.7\linewidth}
\caption{A task structure for \textsf{CHECK BORROWER} and the PSM schema}
\label{hydra.loan.fig}
\end{figure}

In Hydra \cite{PhdThesis:92:Hofstede:DataMod}, behavioural aspects are defined
in process models, task structures
\cite{Report:91:Hofstede:TaskAlg}, 
at all levels of decomposition.
The data modelling technique, PSM
\cite{Report:91:Hofstede:PSM} - an ORM variant, 
also provides decomposition for object types (schema object types decompose
into schemas). Process-centric integration is applied at at all levels of 
decomposition. A conceptual specification language, LISA-D 
\cite{Report:91:Hofstede:LISA-D},
is used to express detailed process specifications and database constraints.
Figure~\ref{hydra.loan.fig} illustrates a decomposed task structure for
\textsf{CHECK BORROWER} and the associated PSM schema.

The contrast with the BNM model for the process model (Figure~\ref{kung.fig}) is
striking. The omission of state-based concerns simplifies considerably the
process model. At the same time, sequence (arrows), decisions (circles) and 
synchronisations (not illustrated) provide the same expressive power as Petri 
net based approaches (in
\cite{Report:91:Hofstede:TaskAlg},
a translation from Petri nets to Task Structures in provided). Of course, more 
complicated task structures are possible than the one bourne out by
Figure~\ref{kung.fig}. 

Task pre- and post-conditions and decisions are specified in 
LISA-D. For example, the LISA-D predicate for the outgoing positive arc
\textsf{y}) of BC1 is:
\begin{eqnarray*}
& & \mbox{borrower } B \mbox{ loans item due-on date} > \mbox{\$TODAY} \\
& & \mathsf{OR} \\
& & \mbox{borrower requires re-registration on date} > \mbox{\$TODAY} +
\mbox{period related-to item } I
\end{eqnarray*}
LISA-D expressions are formulated against the database using an attached
schema. A schema is defined for the highest level task which is visible
to all its decompositions. Buffers (not illustrated) and variables (B, I 
and \$TODAY) may also be defined for temporary storage during task 
processing. A schema type is attached to each buffer. 

A positive outcome results in a termination of
execution returning control to the calling task whereas a negative outcome results
in the \textsf{reject loan request} being executed. Recall (from the BNM example
that this should send a message through the \textsf{rejected request} data flow.
Hydra does not have explicit data flows and so this cannot be specified. Note   
LISA-D's expressiveness does warrant the introduction of a DATE function since
the addition operator is domain sensitive.

The chief strenght of Hydra is its expressive power and formal foundation, both
exceeding that of previously assessed techniques.
The formal semantics for Task Structures are described in Process Algebra  
\cite{Report:91:Hofstede:TaskAlg}. 
A high comprehensibility, particularly for behavioural process modelling is also
clear. A major drawback is business suitability which results from the
previously discussed problem of generality. This together with the lack of
data flow (including messaging) and data store constructs raises uncertainty
about its effectiveness at the business level. 

\subsection{Business-oriented modelling}
\label{bus-model}

As the name suggests, business-orientation does not displace the essential
nature of conceptual modelling, but rather that conceptual modelling is oriented
towards an increased conceptualisation of business aspects. This shifts the 
general (and compromised) suitability of classical techniques, to one which 
is specialised for business domains. Given the lack of a standard inter-subjective
world-view, the strategies for business-orientation vary significantly. 

Earlier attempts at business-orientation focussed on providing some
aspects of organisational processing structure, so that an
\EM{organisational embedding}{\footnote{This name was chosen as a result 
of communication with members 
of IFIP WG 8.1 Task Group FRISCO, a body commissioned to create a FRamework of 
Information Systems COncepts.}}
of conceptual models into different business world-views is possible. 
\cite{Article:85:Richter:PetNets} 
was among the first to use organisational concepts such as actors 
and the cooperative processing of Petri nets to capture the semantics of 
business processes. Earlier attempts at office modelling were based on
completely process-centric techniques, e.g.\
\cite{PhdThesis:89:Bots:ProblemSolving,%
Article:91:Dur:TaskActor}.
The popularity of object concept has resulted in adaption of OOA/D techniques.
In IOOM~\cite{Article:95:Antonellis:OOA}
for example, a purely state-centric approach is proposed using the office
modelling concepts from the
ESPRIT project OSSAD. Object types are defined for business resource
(actors, data or documents) and business processes, and roles
define their permissable states and constraints.

The use of business (or enterprise) models as contexts for conceptual models
provide a particular world-view.
Typically \EM{business plans} 
\cite{Book:85:Davis:MIS}
describe an organisation's operational and
strategic structure through qualitative descriptions of mission, goals, 
objectives, critical success factors, market sectors, competitive 
and quality management strategies). These qualify the business 
services provided and organisational processing structure designed to 
carry out the business services, often occupying significantly sized
documents.
Business models abstract from this detail, describing concepts
such as goal-rooted organisational units, business services, activities, 
tasks, 
actors and actor roles (for processes) and resources (information and 
material). 

\begin{figure}[htb]
\centering
\epsfig{figure=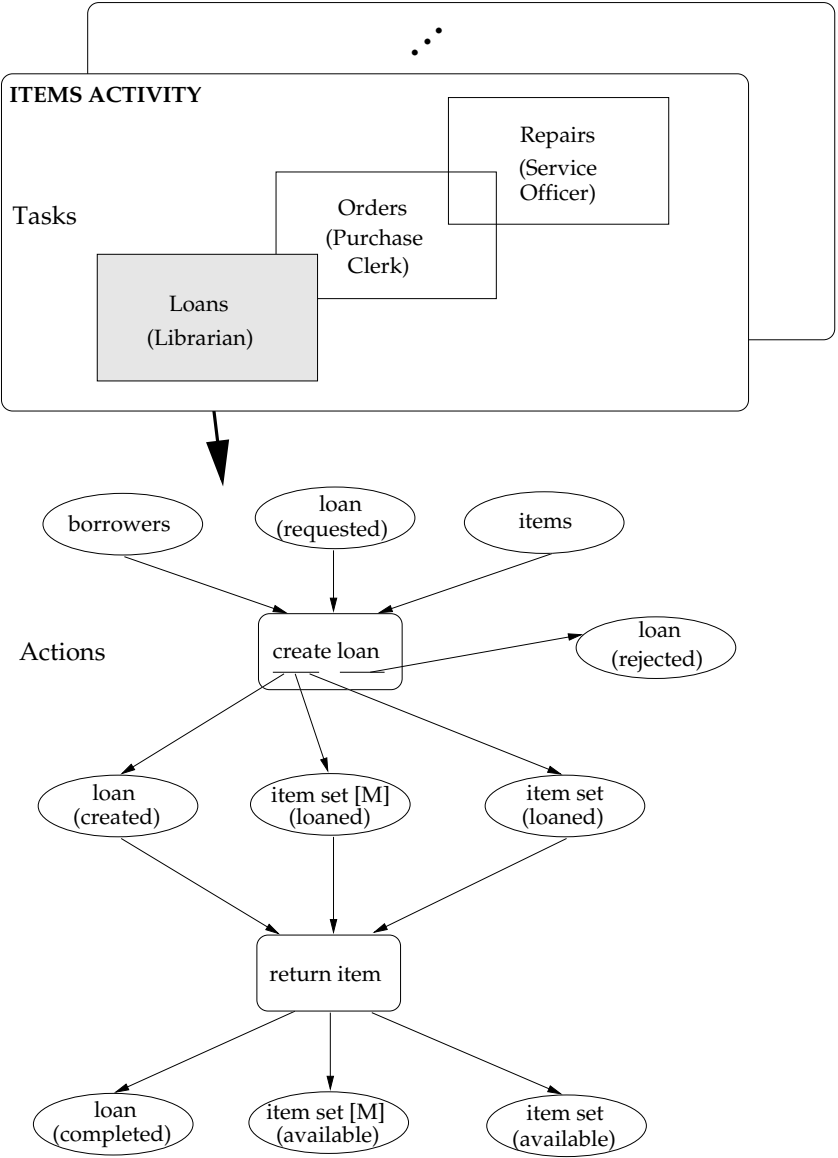,width=0.7\linewidth}
\caption{Operational business modelling for library example in [Ram94]} 
\label{guus.fig}
\end{figure}

\cite{PhdThesis:93:Ramackers:BehaviourMod} provides an integrated 
executable specification framework both for business and IS 
modelling. At the strategic level of the business model, the problem 
solving for business services is carried out. At the operational level, 
concepts exist for activities and tasks, and information, material and actor 
resources. These are anchored into organisational units. Tasks are 
undertaken by a (primary) actor and possibly other actors. They consist 
of a partial order of elementary actions. An action is undertaken by a 
(primary) actor and possibly other actors. Actions apply to 
information/material objects.  Figure~\ref{guus.fig} illustrates a 
part of an (operational) business model for the library domain.
 
Figure~\ref{guus.fig} describes the action processing for the 
\textsf{Items} activity's \textsf{Loans} task which is 
undertaken by the \textsf{Librarian} actor. Task modelling is 
CP-net based. The semantics of each action are as follows: an action 
is triggered when the precondition of all the input resources (actors and objects) 
being in the right state and a business rule (optional) is satisfied, 
the postcondition of resources being moved into the required states results. 
\textsf{create loan} is a composition for two exclusive actions which 
accept or reject the \textsf{loan}. A rule language augments action 
specifications although it is not as expressive as that in Hydra. In the 
example for instance, it is not possible to de-reference objects, 
\textsf{items}, within complex objects, \textsf{loan}. The business
model is effectively developed through a process-centric integration, featuring
the decomposition of an organisational processing structure. 
A number of views may be projected from the business model: data, process 
(note, not conceptually equivalent 
to structured process models), behaviour, object-based (with actions as methods), 
and the information systems in the broader sense (ISN) 
and information systems in the narrower sense (ISN) views motivated 
in~\cite{Article:92:VerrijnStuart:InformationPlanning}.

The IS level is object-based and therefore adopts state-centric integration. 
Those object types which are selected to be computerised are refined at the
IS level (associated actions having a computerised co-actor are defined 
as methods). The behavioural aspects of object types are defined through event 
precedences which trigger state transitions or perform retrievals. Event 
processing  is also CP-net based with an ECA language. Three categories of 
object types are defined. Domain object types are normal IS object types, refined 
from computerised business information object types. View object types aggregate 
domain object types at human-to-computer interaction points (HCI) points which 
are derived through actions involving human and computerised co-actors. View object 
types trigger events on domain object types, therefore classifying application 
transactions. User object types simulate users, triggering events 
on view object types. 
Like the business level, a number of views may be projected: data, process 
(derived through inter-object event triggering, but not equivalent to 
structured process models), object-based, analysis and design.

Of contributory significance 
\cite{PhdThesis:93:Ramackers:BehaviourMod} 
demonstrates that through the increased conceptualisation of business 
aspects, increased conceptualisation at the IS level is possible. As a
result, application design mapping - an area of difficulty traditionally -
is described. That is, through domain object 
type refinement, a business model is mapped to a conceptual IS model while 
through both view and user object type refinements, a business model 
is mapped to an external IS model. Two problems, both related 
to the business modelling technique, exist. The first is that the particular 
organisational processing structure is likely to restrict 
the suitability of the technique for any given business domain. The second, 
and more serious, is the basic treatment of an organisation's interaction 
with its external environment. Specifically, the interaction with the clients 
and other organisations is understood through the internal execution of 
tasks. Tasks form the only context from which actions are triggered.
Events, including external events, are modelled at the IS level only.

To provide an abstraction from organisation processing structures, a new
class of techniques, sometimes referred to as communication-based techniques, 
has been spawned  by the work of
\cite{Article:80:Flores:ComMod}
based on the speech-act theory of
\cite{Book:69:Searle:ComMod}.
The speech-act revisions of
\cite{Book:84:Habermas:ComMod}
have been adopted by more recent such techniques, e.g.\
the Actor-Bank-Channel communication modelling technique
\cite{Article:94:Dietz:TransModelling} of the DEMO method. 
In it, the pattern of performative (state changing) conversations between
actors is used to derive interaction structures for communication units 
known as essential{\footnote{In DEMO the term essential qualifies 
non-computerisable processing despite the possibility of computerised 
support.}} transactions. That is: the request for something, say a business 
service, results in an actagenic (or an action planning) conversation; followed 
by an essential action; and finally a factagenic (or fact generating) 
conversation, stating the results of the action. The actor 
initiating the actagenic conversation, i.e. initiator, is the same as 
the one terminating the factagenic conversation, i.e. executor. Conversely, 
the actor terminating the actagenic conversation is the same as the one 
initiating the factagenic conversation. For the execution of essential actions, 
actors generate plans involving communicative (e.g. information retrieval) 
actions from other actors in order to fulfill the essential action.

The same interaction structure applies to external transaction types, i.e.
the initiator and the executor are external actors (in the environment),
internal transaction types, i.e. the initiator and the executor are internal 
actors (in the business domain), and interface transaction types, i.e. the 
initiator is an external actor and the executor is an internal actor. A 
communication model is developed by building different transaction types 
into an interaction structure and including interstriction details (external 
data sources). A behaviour model is developed through the definition of 
execution and communication rules for performative conversations carried 
out by each actor. Significantly, different behaviour specifications may
apply to different actors for the same object types.

The comparitive advantage of DEMO lies in its actor-centric basis for
business process redesign. This formalises the otherwise ad-hoc 
socio-technical mechanisms in IS methods. Clearly, its most benefitial 
application lies in
``fuzzy'' business processing where the execution paths are uncertain,
infrequent and highly fluctuating - i.e. ones which are difficult to
document in operational procedure. This is well-recognised as a
domain for CSCW.
The benefit for the class of business processing addressed in this paper,
given its overhead remains uncertain. Moreover, its impact at the detailed
design level seems limited (compared to
\cite{PhdThesis:93:Ramackers:BehaviourMod}).
 
\subsection{Summary}
\label{survey-summary}

In this section, the findings of the assessment are generalised for the
paradigms: structured process modelling, object-oriented modelling, 
behavioural-process modelling and business-oriented modelling.

\subsubsection*{Structured process modelling}
Structured process modelling techniques provide a top-down process 
analysis where data flows and stores provide a process-centric integration 
with data modelling techniques. The simple and general set of 
concepts together with the feature of decomposition are applicable for 
different 
types of domains. At higher levels of analysis, models such as DFDs are easy 
to 
comprehend and can be adapted for business domains. A precise definition of 
what
aspects of organisational processing structure should be modelled and how, 
is not available. That only basic aspects of business processes can be modelled
is indicative of an insufficient business suitability. At the lowest level 
where 
the focus is on detailed design, pseudo-code is used to specify processes.
The loosely-coupled integration structure, however, restricts expressive power.
In general, the procedural style of conceptualisation forces a more 
imperative and less declaritive approach to specifications (than in behavioural
process modelling techniques). Also process execution dependencies are not
considered other than the process sequencing implicit in module design 
mapping. 
A major defficiency is the lack of a formal foundation, allowing ambiguities
and inaccuracies in process models.  

\subsubsection*{Object-oriented modelling}
OOA/D techniques tightly-couple object structure and
behaviour into an object model, citing as a motivation the ``natural'' occurence
of objects in most types of domains. A state-centric integration is advocated
principally, whereby an easily comprehensive object lifecycle (typically FSMs) is
defined for each object type. In domains where object interaction is high,
as tends to be the case in business domains, higher level contexts through
state-centric (e.g.\ user cases and scenarios) or process-centric models
(e.g.\ DFDs), are required. These extensions further necessitate an
overall formal foundation, however little evidence of formal semantics was 
found in the surveyed OOA/D techniques. The resolution of difficult design 
decisions such as complex object aggregation and method classification through 
higher level models is therefore intuitive. Again, generality inhibits 
business suitability, although a large class of business rules can be specified 
through object lifecycles. In this regard, a conceptual specification language 
is required to augment ECA specifications. 

\subsubsection*{Behavioural process modelling}
The behavioural aspect of process models fundamentally concerns execution
dependencies between processes, i.e. control flow. This includes the sequence,
repetition, parallelism and synchronisation of processes. Despite generality,
these features allow greater semantics of business processes to be captured.
Traditionally, given their precise and graphically communicable operational
semantics, Petri nets have been used to develop behavioural process modelling 
techniques. As such, model validation through execution is also possible. The 
inclusion of states in process models - note, still process-centric -
provides a declaritive approach to process specifications, and therefore to a 
further class of business rules. However, the impact on the graphical 
representation is problematic, with process models becoming cluttered for even 
basic specifications. Strategies to alleviate this include the use of 
abstraction/decomposition and the use of conceptual specification languages for 
detailed specifications. The use of formalisms like Petri nets does not 
guarantee 
a formal foundation; formal semantics still have to be defined. Algebraic 
systems (such as Process Algebra) provide an alternate mechanism for the 
definition of formal semantics.

\subsubsection*{Business-oriented modelling}
Business-oriented techniques specialise conceptual modelling so that a
precisely defined business suitability is achieved. For this, an 
alignment of concepts with organisational processing structure and a provision 
of colloborative processing are required. 
Unless otherwise addressed, the quality of conceptualisation, expressive power, 
comprehensibility and formal foundation are inherited from the ``underlying''
classical techniques. 
Some techniques cater for a divergent 
set of business world-views through a basic, but not necessarily complete set of 
business constructs. Other techniques prescibe a particular world-view - 
a business 
model. In integrated specification environments, IS models are refined from business 
models and so the formal semantics of the IS model should relate closely to that 
of the business model. Also, the structural dependency of IS models on business 
models can occur, therefore violating the Conceptualisation principle. 
Communication-based techniques focus on organisational communication (a speech-act 
synthesis of actor communication). This leads to a more essential insight into 
business process redesign.

\section{Business suitability principles}
\label{new-principles}

In this section, a diagnosis of the business suitability of
workflow modelling is presented. Recall from section~\ref{intro}, attention
is confined to mission-critical business processing which is amenable to
strict operational procedure. Both business services and business processes
are essential components of this business processing. Although highly
inter-related, they require (as diagnosed) a careful modelling distinction. 

A discussion of the common problems follows, together with an elicitation of
business suitability to alleviate these. The list of principles are:
\begin{itemize}
\item Organisational embedding (section~\ref{embed})
\item Scenario validation (section~\ref{validate})
\item Service information hiding (section~\ref{infohiding})
\item Cognitive sufficiency (section~\ref{combine})
\item Execution resilience (section~\ref{errorhandling})
\end{itemize}

\subsection{Organisational embedding}
\label{embed}

The first problem relates to the relationship between business models and 
conceptual models. On the one hand, the adaption of classical techniques 
with some business-oriented constructs leaves room for an arbitrary 
relationship.
On the other hand, the incorporation of business models might well lead to the 
situation where organisation processing structure prescribes the essential 
structure of the conceptual model; e.g. when an IS model is decomposed 
hierarchically 
from a business model. This, of course, is a violation of the 
Conceptualisation 
Principle. Moreover, it may lead to an inflexibility in IS design 

In  
\cite{PhdThesis:93:Ramackers:BehaviourMod}
for example, it was observed that computerised task actions
are mapped directly into object methods. While an IS design specification
may be decomposed legitimately in this fashion, there are other situations
where IS modules are designed from a composition of tasks, possibly drawn from 
multiple organisations. The point is that such an abstraction should be reflected
at the IS level, and not the business level since no such (single) business task
exists. The task composition serves an IS design purpose only.  

To avoid these problems, the following principle is proposed:

\Principle{\Embed}{  
     A technique should embed all concepts in a conceptual model, 
     directly or indirectly, but without redundancy, into organisational 
     elements.
}

Note, the principle does not prescribe the use of business models. Rather,
it states that any modelling concepts be backtracked to organisational
elements (whatever form of their definition).
In above example, the dependency of IS models on business models is replaced by
an inter-dependency. That is, IS and business models are permitted their
own modelling autonomies, and a consistency between the two is required. The
consistency does not preclude mechanisms such as networked decompositions.
Therefore, business task compositions are permitted for IS process design. 

\subsection{Scenario validation}
\label{validate}

The second problem stems from the observation that all techniques deal, at best,
only partially with the validation of models. Validation is concerned with
ensuring that a conceptual model is indeed a model of a business domain. Beyond
the validation of partial models, no support was evident for an organisationally
embedded validation mechanism which ``cuts across'' the partial models,
drawing their concepts into a unified cognition.
Both the suitability and comprehensibility of a technique are important
factors in model validation, and it was evident in the survey that
different integration strategies offered different advantages.

In classical techniques it was observed that triggering sources
such as external events are used to trace the execution paths in models.
In process-oriented techniques the resultant process execution sequence
is determined implicitly in structural process models and explicitly in
behavioural process models. The validation of execution paths is also advanced
in behaviour process models through synchronisation and decision constructs.
In OOA/D techniques, process sequences apply when object interactivity is high
while object state sequences (in accordance with object lifecycles) apply when
object interactivity is low. In business-oriented techniques, it was quite
clear that organisational processing structures or organisation communication
provide focal units for business processing.

A common theme for model validation is the notion of event. At discrete points
within a timeline, it can be seen that an event triggers actors to execute
business processes, through which information may be accessed, and further
events, possibly invoking some processing in the environment (i.e. outside
the business domain), may result. Intrinsic to the initial event is some
defined purpose which motivates an execution path all the way to the final
event. The final event signifies the (logical) termination of processing;
i.e.\ representing the organisation's recognition that no further processing
should proceed. Such an execution path is referred to as a \EM{scenario}. For
the purposes of model validation therefore, the following principle is proposed:

\Principle{\Validate}{
      A technique should provide an explicit notion of scenario
      for model validation.
}

Note, the principle does not exclude other forms of model validation, e.g.\
the validation of constraints in a data model. Rather, it requires that
scenarios, amongst other possible forms of validation, be supported by a
technique. At the same time, is should be pointed out that since a 
scenario provides a systematic means by which execution paths are traced, it 
should lead to different parts of conceptual submodels, from
which further validation may then follow. Hence, scenarios should provide 
a more effective completeness check than partial model checks.

Of issue is the organisational embedding of a scenario. An obvious choice
is the notion of a \EM{business transaction}, drawn from a Macroeconomics
perspective of organisations
\cite{Book:91:Boyes:BusMod}.
Business transactions concern the exchange of (goods and) services using
business transactions as the fundamental (accounting) unit. Clearly, the
determination of a business transaction cost requires an understanding of
the processing undertaken. It may be simple involving a low interactivity:
a small set events and business processes. Or it may be complex involving a
high-interactivity over a ``long'' temporal duration: a large set of events
and business processes, typically involving multiple organisations.

At a first glance, it may seem that the choice of business service over
business transaction as a scenario is arbitrary. Certainly, both terms tend to
be used interchangeably in IS research.  A cause-effect distinction however,
leads to the preference of business transactions. That is, business services
are a cause reflecting, more than anything else, client requirements of the
organisation whereas business transactions are the effect; in effect, the
scenario which results. In this sense, the concept of business transaction
as an organisational artefact may be unified with the concept of workflow as
an IS design and implementation artefact, i.e.\ \EM{business transaction
workflow}. In stating this, it should be understood that that class of business
transactions involving uncertain, indeed ``fuzzy'' execution paths, has been
excluded from the scope of this paper (section~\ref{intro}).

\subsection{Service information hiding}
\label{infohiding}

Following from the preceding discussion, the third problem lies in the 
relationship between events, business processes and business 
services. Recall, events trigger process execution for some ultimate
intention. Moreover, a number of events may be related to the same 
intention. In a business sense, intentions are denoted by business services. 

In classical modelling techniques, no explicit service concept exists.
Business services are therefore dealt with in an arbitrary fashion.
In process modelling
techniques for instance, services are qualified (informally) in event 
descriptions and therefore directly trigger processes (see examples in 
\cite{Book:89:Yourdon:Analysis}).
Similarly in OOA/D techniques, events directly trigger methods in object 
types (recall Figure~\ref{shlaer.fig}). 
Since business-oriented techniques adapt classical techniques, their treatment 
of business services is also arbitrary. For example in
\cite{PhdThesis:93:Ramackers:BehaviourMod},
business services (functions) are related to business processes (tasks),
but this is no more than a reference; i.e. triggering relates directly to
business processes. 

The common problem in these approaches is the \EM{direct} triggering
of processes (or methods) given the context of triggering. From the point 
of view of the environment or from different parts of an organisation, the 
actual business processes triggered for some business service request are 
inconsequential for the formulation of the request. That is, the request is 
issued for a business service and as a result some internal mechanism is used 
to determine what action to take. The implication is that when processes are 
reengineered, the actual request is not affected.

It may be recognised that the above paragraph reflects the \EM{Information 
Hiding} principle for (software) module design which requires that internal 
aspects of modules be insulated from other modules. When applied to a 
technique's treatment of business service triggering, it can be seen that a 
service concept should explicitly be supported to coordinate the execution of 
number of processes and insulate them from service requests. Or in a more general 
form, the following principle results:

\Principle{\InfoHiding}{
     A technique should allow the formulation of service requests to
     be independent of their actual processing.
}

\subsection{Cognitive sufficiency}
\label{combine}

The fourth problem is fairly general in its diagnosis. It addresses the missing
constructs in techniques which yield collectively a sufficient 
\EM{cognition}{\footnote{A pragmatic rather than a theoretical diagnosis is 
presented. This is based on the comparative differences in the assessed 
techniques and their advantages on model cognition.}}
of a business processing model, i.e.\ business transaction model. 
Fundamental areas 
of variance are discussed below.

\subsection*{Process modelling cognition}
The process modelling in most techniques was observed to incorporate essentially
either structural (data flows) or behavioural (control flows) aspects, but not 
both.
In BNM, DFDs and PrT-nets are both used, but at different levels of abstraction.
In 
\cite{PhdThesis:93:Ramackers:BehaviourMod}, 
a process model view is derived from process behaviour, and so it is not 
structural. Specifically, the data flows are a set of attribute values rather 
than identifiable \EM{containers} of those values, while data stores are 
object 
states rather than identifiable repositories of data.
In Hydra, a `middle ground' is apparent where process behaviour is combined 
with some structural aspects: process decomposition and a notion of data stores 
(temporary storage).
At the other extreme, in Structured Analysis, process behaviour is altogether 
omitted from DFDs except for process sequencing (only) defined as part of 
software design mapping.

Structural and behavioural aspects of process models serve distinctly
different aspects of analysis.
Data flows, stores and transformations from the structural aspect lead to an 
understanding of data dependencies, from which business processes and data 
stores may be redesigned. As examples, common patterns of data flow in different 
process types indicate the need for process reclassification, while dependencies 
on a number of data stores for a data transformation function indicate the need for 
data relocation. Process sequence, repetition, parallelism and synchronisation
from the behavioural aspect, on the other hand, lead to an understanding of
process dependencies from which the problems of deadlock, livelock and starvation 
may be identified. As a result, execution paths may be optimised or processes
themselves may be reclassified.
Interestingly, the strategies defined in
\cite{Article:90:Hammer:BPR} indicate that a combination of both -
simultaneously - is required for business processing reengineering. This
indicates that both are required to sufficiently capture and understand
a business transaction's semantics.

\subsection*{HCI}
Another area of cognitive insufficiency is the ad-hoc conceptualisation of 
HCIs (recall human to computer interactions). For HCI, two issues are apparent 
\cite{Report:89:Brinkkemper:DialogueSpec}, 
namely the point (process) at which a HCI is required and the \EM{dialogue} which 
determines the screen/form design.
Under classical analysis, as evident from the discussions of process
and object modelling techniques, HCIs are determined as part of systems design
where application services, which typically encapsulate one or more
screens/forms, are determined intuitively from conceptual models. This situation, 
as remarked more than once, leads to a waterfall since the backtracking of 
specifications may be difficult. In 
\cite{PhdThesis:93:Ramackers:BehaviourMod},
it was seen that business model semantics, in this case the actor types
involved, can be used to derive HCI points.
\cite{Article:94:Dietz:TransModelling}
synthesises actor-actor communication so that dialogues go beyond HCI
where it is argued, systems design including HCI, may 
be better understood.

\subsection*{Temporal aspects}
Finally, the treatment of temporal aspects is partial despite the (obvious)
impact of time in business domains. In Structured Analysis and other DFD
techniques, a ``clock'' symbol is introduced to indicate time triggers
on processes, e.g.\ end of the week. This however is only a comprehensibility
feature. In
\cite{PhdThesis:93:Ramackers:BehaviourMod},
time triggers are incorporated into the event specification language for
object types, thereby extending the concern to expressive power. Further
to expressive power, the need for time functions and variables was indicated 
in the discussions of BNM and Hydra (recall the preconditions of \textsf{BC1}). A 
survey of techniques (including TEMPORA) dealing with time
\cite{Article:91:Theodoulidis:TimeModel},
demonstrates the incorporation of time on action preconditions, and relatedly
dynamic constraints on object type specifications.

It is clear that these incorporations of temporal aspects relate to the 
specification of process preconditions. Of course, time should also be 
important in postconditions, e.g.\ as motivated by event scheduling 
requirements in 
\cite{PhdThesis:90:Brinkkemper:Formalisation}. 
This of course further qualifies process dependency since a number of 
processes may be required to execute at the at some time simultaneously 
(parallelism) or within a time duration of each other (sequence). 
Also a process may execute repeatedly within a certain time period 
(repetition), until some condition is satisfied. It is interesting to note that 
some model-based formal specification languages e.g.\ 
\cite{Article:86:Zave:FormalLang},
specification languages e.g.\ Real-Time Process Algebra
\cite{Article:91:Baeten:RealTimePA},
and Petri net based approaches e.g.\ ExSpect
\cite{Article:89:Hee:Exspect},
all address temporal aspects in process specifications.
Indeed, the traditional domain for this consideration has been in real-time
systems. The need of temporal specification, is nonetheless, evident in
business domains. For example in law courts, matters are scheduled and adjourned
at designated times, and in doing so, the availability of required 
documentation is requested (to outside organisations and parties) within 
a certain duration prior to the court hearing. 

\Principle{\Combine}{
     A technique should provide a sufficient cognition of a model 
     such that the need for fundamental business process execution
     assumptions is eliminated.
}

\subsection{Execution resilience}
\label{errorhandling}

The fifth problem relates to error handling at the conceptual 
level. Of course at the conceptual level, database constraints and 
process pre- and post-conditions define an error free IS state.
Moreover, specifications may be verified to eliminate erroneous specifications. 
However, \EM{operational} errors can still occur beyond the control of 
an IS. For example, in a library domain, the existence of an overdue item (over 
different categories of duration) is an operational error(s) caused by a 
borrower. Similarly, a missing enrollment confirmation in the required time 
since an initial enrollment, is caused by a student. Finally, a system crash is 
one of several examples of operational errors resulting in non-deterministic 
processing failures. 

A sound conceptualisation of business transaction semantics requires the 
the inclusion of operational error handling. However, no evidence in the 
assessed techniques indicated an explicit consideration of this.
In most cases, an expressive power is available for dealing 
with errors like the first two examples. In process-centric techniques for
example, error handling processes may be defined to scan a database for 
problematic object states (overdue item, unconfirmed enrollment). In 
state-centric techniques, an error handling process may be fired as a result 
of an object's transition into a problematic state.

The treatment of non-deterministic failures has recently become the 
subject of workflow implementation specifications. In particular, the 
traditional transaction model (more recently and comprehensively described in 
\cite{Book:93:Gray:TransMgmt})
with its ACID properties (atomicity, consistency, 
isolation and consistency) has been extended for workflow execution semantics. 
Under the traditional model, a transaction binds a set of database 
operations into an atomic unit of execution. Following the requirement of 
\EM{failure atomicity}, a transaction's changes to a database(s) are 
\EM{committed} if the execution is successful or \EM{rolled-back} if not. 
Following the requirement of \EM{execution-atomicity}, the concurrent 
execution of transactions should have the same effect as if they were executed 
in a \EM{serialisable} order. 

Workflows are more complex structures than traditional transactions, and it is 
unacceptable that the failure of any one of its tasks results in the rollback 
of the entire workflow. The application of transaction models to workflows,
typically through a relaxation of ACID, are surveyed in
\cite{Book:94:Kim:TransMgmt} (pp.596-598). Under an ACID relaxation, 
a failure atomicity is defined for each task. This involves either an \EM{undo}, 
possibly through the execution of some other task, i.e. a \EM{compensation}, or a 
\EM{redo}, possibly through a contingent task execution. Committed tasks are
therefore not re-executed. Committed tasks, however, may update data objects and 
release locks, allowing other tasks and transactions to ``see'' the updates. Hence 
the need for compensations, which either logically undo the updates or provide 
some notification of update invalidity. Of course, compensating tasks should not, 
themselves, fail.

The lack of explicit support for operational error handling is indicative of
the traditional IS development approach which prefers this treatment at the
the implementation level. A certain amount of this detail is, afterall,
implementation-oriented (e.g.\ checking for a DBMS-defined deadlock error 
code after an update to determine whether a retry should be issued). Yet it 
should be recognised that operational error handling is an inherent part of 
a business transaction's semantics. Where the previously described 
conceptualisation aspects relate to the \EM{\Combine} of
normal execution semantics, the conceptualisation of operational error
handling relates to \EM{execution resilience} semantics. As repeatedly 
discussed,
the partial conceptualisation of some aspects and not others, can distort
the overall understanding of the business transaction semantics.

Significantly, the support of model execution in a large number of techniques, 
means that execution resilience can actually be validated and ``tested'' at the 
conceptual level. Given the disjointness (and complexity) of operational
error handling - best presented as a separate layer to preserve a model's 
general cognition - model 
execution is considered critical. Therefore the following principle results: 

\Principle{\ErrorHandling}{
   A technique should support the handling of
   operational errors, so that a resilient execution of the
   conceptual model results.
}

Ultimately, such support addresses a workflow's execution resilience. This of
course does not guarantee execution resilience since operational errors, 
e.g. crashes, can (and do) happen.

\section{Epilogue}
\label{epilogue}

\subsection*{General problem}
Conceptual modelling techniques haven proven to be invaluable
for early and critical phases of analysis and design. When
``bundled'' into well-formed IS methods, a navigation from
organisational analysis to implementation specifications is
possible. For an essential problem-solving insight into
different aspects of business domains, techniques should
demonstrate a conceptualisation, expressive power, comprehensibility, 
formal foundation and business suitability. In the last of these 
aspects, the absence of a universal organisational theory has meant 
that techniques can only increase in technical effectiveness as 
newer insights are obtained from practical experience, and as the 
functionality of IS platforms expands. This makes it difficult to 
assess the adequacy of techniques to support a sound conceptualisation 
of business domains.

\subsection*{Focus of paper}
In this paper, the business suitability of workflow modelling
was diagnosed for that class of business processing which is
mission-critical in nature and which is amenable to strict operational 
procedure. The approach taken was two-fold. First, the general 
capabilities of techniques characteristic of well-known paradigms were 
assessed using the general requirements. Second, the extent to which
workflows scale out to a business suitability - ``constructed'' from the 
business processing inter-relationship of business service \EM{and} business 
processes - this combination providing a new insight in our opinion - was 
diagnosed. As a result, five new business suitability principles were 
formulated.

\subsection*{General assessment}
The findings of the general assessment were as follows. 
Classical techniques, i.e. process (structured and behavioural) and 
object modelling, offer general concepts and features which are applicable to 
several types of domains. Where they differ is in the cognitive dependencies of
their integration strategies. Process-centricity emphasises data flow
transformations (structured) and process control dependencies (behavioural).
Both are useful for describing the flow of business processing.
State-centricity emphasises data object states and the transitions between
them. This is useful for focusing event (signaling/messaging) and their impacts 
on business processing. Since an object lifecycle deals only with one object type, 
a number of OOA/D techniques also incorporate some form of process-centricity.
In general, the need for a trade-off between comprehensibility and 
expressive power was conspicuous throughout the assessment. Techniques biasing
the former (e.g. DFDs) tend to be weak in the latter (e.g. pseudo-code), 
conversely those biasing the latter (e.g. Petri net based approaches)
tend to be weak in the former (e.g. cluttered diagrams). Also an adequate
formal foundation, in particular formal semantics, was notably absent in most
techniques. This raises ambiguities in the application of techniques, and
in turn makes it impossible to prove certain properties about their models. In this
regard, the absence of correctness checking in workflow specifications (noted in
section~\ref{intro}) is striking. 

Business-oriented modelling adapt combinations of classical techniques within
the context of some inter-subjective business world-view. This may be partial,
i.e.\ the incorporation of some (key) business concepts, or it may be complete, 
i.e.\ the incorporation of a business model. The use of essential business
modelling concepts such as business services, processes (of various
specialisations), actors and actor roles, organisational units and resources
(material and information) has become popular, particularly through integrated 
specification environments such as CAiSE tools and requirements engineering 
methods. Several approaches adopt a decomposition of organisational processing 
structure. In so doing, a more automatic application design 
mapping is demonstrated compared to classical techniques. Recently proposed
communication-based techniques abstract from organisational processing structure
and analyse human speech-acts as a basis for understanding business processing
(amongst other forms of processing). This allows more detailed aspects of 
workflow
interaction to be understood, however the extent of its applicability to the
class of business processing we considered was not clear.

\subsection*{Business suitability principles}
The detailed diagnosis of business suitability led to the identification of
the following problems, for which associated principles were defined:
\begin{itemize}
\item A hierarchical decomposition from a business model to an IS model
can result in a violation of the Conceptualisation Principle whereby
an essential structure of the IS model is prescribed. For example, a
hierarchical decomposition does not allow IS processes to be composed out of
distinct organisational processing structures, say from more than one 
organisation. To provide techniques with their own abstraction autonomy, 
hierarchical decomposition is replaced by the \EM{\Embed\ Principle}. It 
simply requires a link of any given IS concept, whether direct or indirect, 
into organisational elements.
As a result, the inter-dependency between the business and IS levels does
not impact the conceptualisation at these levels.
\item Inherent in business processing is a scenario, i.e. an execution 
sequence which occurs over a time duration for some particular intention. 
Scenarios are important since they provide a domain semantic interpretation, 
i.e. validation, of a model. Support for scenarios \EM{per se}, is absent in 
classical techniques, and present in some form, through either organisational 
processing structures and organisation communication, in business-oriented
techniques. The need for explicit support was defined through the 
\EM{\Validate\ Principle}.
The notion of business transactions, recognised within Macroeconomics theory, 
further qualified the business service and business process inter-relationship 
as a scenario. That is, the triggering of a business service for a request, and 
any subsequent triggering, including those on business processes and business 
services resulting in (some defined) satisfaction of the request, constitutes a 
business transaction; an essential scenario concept.
\item Related to the above, all techniques directly or potentially lead
the knowledge of ``internal'' processing required during the formulation of
a service request. This, of course, violates the well-known Information
Hiding Principle. To alleviate this, the \EM{\InfoHiding\ Principle} was defined. 
As such, service requests are completely insulated from subsequent triggering. Put
simply, services coordinate process execution.
\item General omissions were identified which restricted the cognition of a
business transaction model. These included: the support of either structured 
(e.g.
data flow) or behavioural (e.g. control flow) aspects for process modelling,
but not both; the ad-hoc conceptualisation of HCIs; and finally (and surprisingly)
the partial support for temporal aspects (e.g. not considered in process 
postconditions). When not supported by a technique, erroneous assumptions can
result. The \EM{\Combine\ Principle} was defined to ensure that such fundamental 
assumptions are eliminated.
\item Operational errors are those which occur during IS operation, but
outside control of the IS. Although most techniques have an expressive power
to support some types of operational error handling specifications,
and although the handling of
non-deterministic operational failures is considered crucial in workflow specifications,
no explicit support is provided by the techniques. When specified at the 
implementation, level, the overall context of business processing semantics may be 
lost. Furthermore, any model validation through execution is partial, bypassing 
operational errors.
The \EM{\ErrorHandling\ Principle} was defined to ensure that operational errors 
are specified at the conceptual level, so that a model may be ``tested'' for 
execution resilience. This addresses the execution resilience of an IS ultimately.
\end{itemize}

\subsection*{Conclusion}
To conclude, the business suitability principles address an important yet
not often salient aspect of IS conceptual modelling. Of course, we focussed
the diagnosis on a particular aspect of business domains, namely a particular 
class of business processing, characterised in a particular way, and intended for
a particular implementation model - namely that based on the workflow concept.
Undoubtedly, further considerations of this and other aspects of business domains 
are still required. Importantly, our characterisation 
of business processing is well-recognised, and the conceptual level of diagnosis 
imparts relevance to traditional implementation models as well. 
Moreover by describing rather than prescribing the principles, we seek to 
position our work more effectively with other such developments. 
 
The major benefit lies in assessing techniques, whereby immediate attention may be
drawn to the general areas of deficiency we encountered. To overcome these and other 
deficiencies, it should now be clear that enhancing
a technique requires a simultaneous consideration of all the conceptual modelling 
requirements. Given the large number of useful contributions in the field, we advocate 
a synthetic rather than analytic approach to developing techniques. 

Another benefit lies in the insight gained for workflows involving global access
(e.g.\ via the Web), inter-organisational collaboration and reuse of application 
services. A clear separation between business service, business process and
business transaction not only improves the conceptual modelling but may
well have far reaching
implications for workflow specification languages and, ultimately, for 
(middleware) description languages of open distributed architectures. 

Future work will deal with the development of a \EM{kernel} of concepts 
and features for workflow modelling along the lines of the principles.


\begin{scriptsize}
   \BIBLIOGRAPHY{alpha}
\end{scriptsize}

\end{document}